\documentclass[pra, showpacs, twocolumn,preprintnumbers, a4paper]{revtex4}
\usepackage{graphicx}
\usepackage{graphics}
\usepackage{amsmath, amsfonts, amssymb, bm}

\usepackage{amsmath}
\usepackage{verbatim}
\usepackage{bm}

\usepackage{color} 

\begin{document}

\newcommand{\ket}[1]{| #1 \rangle} \newcommand{\bra}[1]{\langle #1 |} \newcommand{\braket}[2]{\langle #1 | #2 \rangle}

\title{Nonlocal fluctuations and control of dimer  entanglement dynamics} 

\author{ Cristian E. Susa and John H. Reina}
\email{jhreina@univalle.edu.co}
\affiliation{Departamento de F\'isica, Universidad del Valle, A.A. 25360, Cali, Colombia
}

\date{\today}

\begin{abstract}
We report on the dissipative dynamics of an entangled, bipartite  interacting system.
  We show how to induce and control  the so-called 
 early stage disentanglement (and  the `delayed' entanglement generation) dynamics 
by means of a driving laser field. 
We demonstrate that some of the features currently associated with pure non-Markovian effects in such entanglement behavior can actually take place in   Markovian  environments if background noise  QED fluctuations are considered. 
We illustrate this for the case of a dimer interacting molecular system
for which emission rates, interaction strength, and  radiative corrections have been previously measured. We also show that even 
in the absence of collective decay mechanisms and qubit-qubit interactions, 
the entanglement still exhibits collapse-revival behavior.  Our results  indicate that  
zero point energy fluctuations should be taken into account when formulating precise entanglement dynamics statements.
\end{abstract}

\pacs{03.65.Ud, 03.67.Mn, 42.50.Fx, 42.50.Lc, 33.50.Dq}


\maketitle
\section{introduction}
Entanglement is arguably the most striking  feature of 
quantum phenomena \cite{mermin}, and  is a crucial physical resource to quantum  computing \cite{bennett}, and quantum communication via  protocols such as 
teleportation \cite{bennett1} and cryptography 
\cite{gisin}. Works on entanglement dynamics  
have recently shown that some initial bipartite entangled states 
 can decay to zero in a finite time much shorter than that of their spontaneous emission 
 \cite{yu1, konrad, dodd, mintert, luis, scala, yu2}. This phenomenon--early stage disentanglement (ESD)-- is a quantum  feature that signals an unusual dissipative dynamics of purely non-local quantum correlations \cite{mermin,bennett1}. It has also recently  been shown  that, in  specific scenarios, it is possible to produce a delayed creation of entanglement, the so-called entanglement sudden birth  (ESB) \cite{ficek, davi07}. These  
entanglement  features have been explored in both the Markovian  \cite{yu4} and Non-Markovian  \cite{bellomo}
regimes in several different contexts \cite{ficek,yu2,davi07,yu4,bellomo,jh1}.  
Even more  recently, there have been reports on the quantification of non-classical correlations of bipartite qubit systems in interaction with non-Markovian reservoirs \cite{amir} by means of the quantum discord \cite{zurek}
dynamics, in contrast to that of entanglement evolution \cite{amir}.

In this work, we show how the ESD and ESB phenomena can be externally  coherently controlled with a laser field. We also show that the entanglement dynamics is permanently affected by the background noise  due to nonlocal vacuum fluctuations, a fact that permanently affects the ESD dynamical profile. Our results give a practical prescription, within the Born-Markov formalism, for the identification and control of ESD, ESB, and collapse-revival phenomena. This shows that   previously identified  entanglement dynamics associated to non-Markovian behavior \cite{bellomo}, can actually take place within a Markovian dynamics description.

Most  of the interacting two-qubit systems can be described by a   Hamiltonian of the type  \cite{hammerer}   $\hat{H}_S=\hat{H}_0+\hat{H}_{12}$,
 where $\hat{H}_0$ denotes the free particle term, 
and 
$
      \hat{H}_{12} =h\big(
      J_{x}\sigma^{(1)}_{x}\otimes\sigma^{(2)}_{x}+
      J_{y}\sigma^{(1)}_{y}\otimes\sigma^{(2)}_{y}+
      J_{z}\sigma^{(1)}_{z}\otimes\sigma^{(2)}_{z}\big)
$,
written in terms of the Pauli matrices, denotes a generic, anisotropic qubit-qubit  interaction, 
and $h$ is Planck's constant. In many physical processes, such as energy transfer in biomolecular systems \cite{jh1,jh1a} or interacting artificial atoms \cite{jh2}, to name but a few,  $J_x=J_y\equiv J_{xy}$, and the strength  ratio $J\equiv J_{xy}/J_{z}$ sets  the type of interaction experienced 
by the coupled qubits, and  defines  the type 
 of entanglement that can be `naturally' \cite{hammerer} generated in the bipartite  system.
 
In this article, we consider the case of dipole-dipole (d-d)  interacting single
molecules of transition frequencies $\nu_{i}$ (molecule $i$) in 
 interaction with the quantized radiation
field, and externally driven by a coherent  laser field  \cite{hettich,jh3}. We denote  by $\ket{0_{i}}$, and $\ket{1_{i}}$, $i=1,2$, the ground and excited state of molecule  $i$, respectively, and hence $\hat{H}_0=-\frac{h}{2}\nu_1\sigma^{(1)}_{z}-\frac{h}{2}\nu_2\sigma^{(2)}_{z}$.
The  two
two-level molecules are separated by the
vector $\mathbf{r}_{12}$ and are characterized by  transition dipole moments
 $ \hat{\boldsymbol{\mu}}_{i}\equiv\bra{0_i}\mathbf{D}_{i}\ket{1_i}$, with dipole operators $\mathbf{D}_{i}$, and spontaneous
emission rates $\Gamma_i$.
The  system interaction Hamiltonian can be written in 
the computational basis of direct product states $\ket{i}\otimes\ket{j}$ $(i, j=0,1)$ as
$\hat{H}_S=\hat{H}_0+\hat{H}_{12}$, where   
$\hat{H}_{12}$ is set by the 
dipole coupled  
molecules of interaction energy  $hV_{12}$ ($J_x=J_y\equiv V_{12}, J_z\equiv 0$).

\section{Dimer  dissipative dynamics}

Evidence of the molecules dipolar 
coupling, and the generation of sub- and super-radiant
states, as well as an effective shift of the doubly excited state, have been experimentally reported in \cite{hettich}. The latter  has been accounted for in $\hat{H}_{12}$ as a phenomenological, effective $\Delta_{e}$-shift of the doubly-excited state $\ket{11}\equiv\ket{1_{1}}\otimes\ket{1_{2}}$, due to the vacuum fluctuations \cite{jh3}. 
The dimer is externally controlled by a coherent driving  field 
which  acts on each of the molecules with a coupling 
amplitude $h\ell_{i}=-\boldsymbol{\mu}_i\cdot \boldsymbol{E}_i$, and a frequency $\omega_{L}$. The light-matter interaction Hamiltonian  $\hat{H}_L=h \ell^{(i)}(\sigma^{(i)}_{-}e^{i\omega_{L}t}+\sigma^{(i)}_{+}e^{-i\omega_{L}t})$, where $\boldsymbol{\mu}_i$ is the $i$-th transition dipole moment  and $\boldsymbol{E}_i$ 
is the amplitude of the coherent driving acting on molecule $i$ (at $\boldsymbol{r}_i$). 

\begin{figure*}[t]
  \centerline{\includegraphics[width=0.9\textwidth]{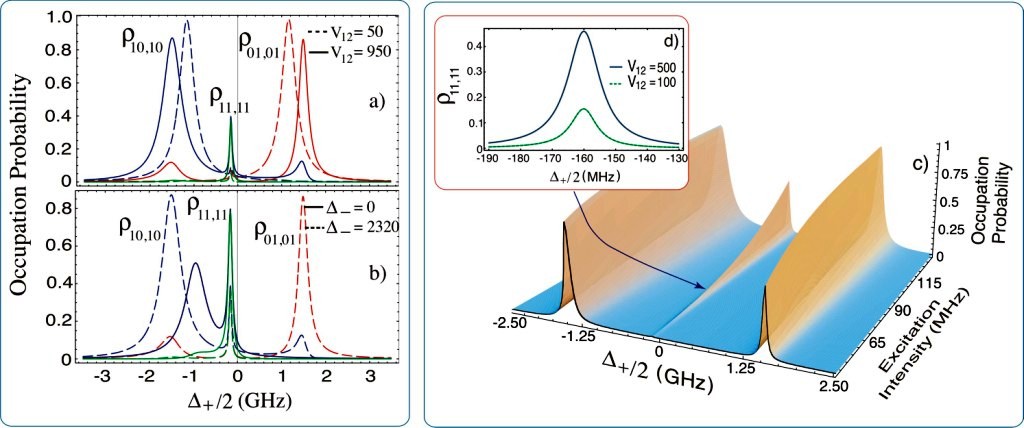}}
  \caption{Steady state occupation probabilities for the dimer system.   The detunings $\Delta_{-}\equiv \nu_{1}-\nu_{2}$, and $\Delta_{+}/2\equiv(\nu_{1}+\nu_{2})/2-\nu_{L}$. a) Populations $\rho_{ij,ij}$ for two different dipolar couplings, $\Delta_-=2320$ MHz. b) Probabilities for a fixed dipolar coupling, $=950$ MHz, and different molecular resonance conditions. In both cases, $\Gamma_1=\Gamma_2=2\pi\times50$ MHz, $\Gamma_{12}=2\pi\times9$ MHz, $\Delta_e=-160$~MHz and $\ell_i=200$~MHz. c) Fluorescence spectrum given in terms of the occupation probabilities, as a function of the laser coupling strength and the laser detuning $\Delta_{+}/2$. In MHz, $\Gamma_{i}=18\pi$, $\Gamma_{12}=9\pi$, $\Delta_-=2320$, and $\Delta_e=-160$. d) Zoom of the two photon (intermediate) resonance for $\ell_i=100$~MHz.}
\label{observable}
\end{figure*}

The dimer  dissipative dynamics is appropriately described within the Born-Markov formalism, with $\hat{H}=\hat{H}_S + \Hat{H}_{L}$, 
by  the quantum master equation
\begin{equation}
\label{master}
\hat{\dot\rho}= -\frac{i}{\hbar} \big[\hat{H},\hat{\rho}\big]  + L(\hat{\rho}) ,
\end{equation}
with  dissipative Lindblad 
super-operator given by  \cite{agarwal}
\begin{eqnarray}
     L(\hat{\rho}) &=&
      -\frac{\Gamma _{1}}{2}\left( {\hat\rho}
      \sigma^{(1)}_{+}\sigma^{(1)}_{-}+\sigma^{(1)}_{+}\sigma^{(1)}_{-}{\hat\rho}
      -2\sigma^{(1)}_{-}{\hat\rho} \sigma^{(1)}_{+}\right) \\
    &  &
 -\frac{\Gamma _{2}}{2}\left( {\hat\rho}
      \sigma^{(2)}_{+}\sigma^{(2)}_{-}+\sigma^{(2)}_{+}\sigma^{(2)}_{-}{\hat\rho}
      -2\sigma^{(2)}_{-}{\hat\rho} \sigma^{(2)}_{+}\right)  \nonumber \\
    &  &
-\frac{\Gamma _{12}}{2}\left( {\hat\rho}
      \sigma^{(1)}_{+}\sigma^{(2)}_{-}+\sigma^{(1)}_{+}\sigma^{(2)}_{-}{\hat\rho}
      -2\sigma^{(1)}_{-}{\hat\rho} \sigma^{(2)}_{+}\right) \nonumber \\
   &  &
 -\frac{\Gamma _{21}}{2}\left( {\hat\rho}
      \sigma^{(2)}_{+}\sigma^{(1)}_{-}+\sigma^{(2)}_{+}\sigma^{(1)}_{-}{\hat\rho}
      -2\sigma^{(2)}_{-}{\hat\rho} \sigma^{(1)}_{+}\right), \nonumber
\label{diss}
\end{eqnarray} 
where $\Gamma_{ii}\equiv \Gamma_i=nA_{12;i}/2=n\omega_i^3\|\boldsymbol{\mu}\|^2/(3\epsilon_0 hc^3)$ are the 
spontaneous emission rates ($A_{12;i}$, $i=1,2$ are the Einstein vacuum coefficients), $n$ is the refraction index 
of the dispersive medium in which the molecules are embedded 
 (for example, a paraterphenyl crystal in \cite{hettich}), 
and $\Gamma_{12}=\Gamma^{\ast}_{21}$ are the collective spontaneous emission 
rates, arising from the coupling between the molecules through the 
vacuum field \cite{agarwal}; $\sigma^{(i)}_{+}=\ket{1_{i}}\bra{0_{i}}$, and $\sigma^{(i)}_{-}=\ket{0_{i}}\bra{1_{i}}$ are the raising 
and lowering Pauli operators acting on  molecule $i$.

In particular, 
in the near field approximation ($r_{12}\ll\lambda_L$) 
the d-d interaction energy is given by  
$hV_{12}= \frac{3h\sqrt{\Gamma_1 \Gamma_2}}{8\pi z^3}\Big[\hat{\boldsymbol{\mu}}_{1}\cdot  \hat{\boldsymbol{\mu}}_{2}
-3(\hat{\boldsymbol{\mu}}_{1}\cdot
      \hat{\mathbf{r}}_{12})
(\hat{\boldsymbol{\mu}}_{2}\cdot
      \hat{\mathbf{r}}_{12})\Big ]$, and $\Gamma_{12}=\sqrt{\Gamma_1 \Gamma_2}\, 
\hat{\boldsymbol{\mu}}_{1}\cdot  \hat{\boldsymbol{\mu}}_{2}
$: the maximum (minimum) interaction  strengths $V_{12}$ are obtained for
parallel (perpendicular) dipole moments  \footnote{In general, 
the 
expressions for the d-d interaction strength and the incoherent 
decay rate $\Gamma_{12}$ are given by 
$
V_{12}=
\mathcal{A}_{12}
\{-[\hat{\boldsymbol{\mu}}_{1}\cdot  \hat{\boldsymbol{\mu}}_{2}
-(\hat{\boldsymbol{\mu}}_{1}\cdot
      \hat{\mathbf{r}}_{12})
(\hat{\boldsymbol{\mu}}_{2}\cdot
      \hat{\mathbf{r}}_{12})] \frac{\cos z}{z} +
 [\hat{\boldsymbol{\mu}}_{1}\cdot  \hat{\boldsymbol{\mu}}_{2}
-3(\hat{\boldsymbol{\mu}}_{1}\cdot
      \hat{\mathbf{r}}_{12})
(\hat{\boldsymbol{\mu}}_{2}\cdot
      \hat{\mathbf{r}}_{12})]
[ \frac{\cos z}{z^{3}}+\frac{\sin z}{z^{2}}]\}
$, and 
$\Gamma_{12}=
\mathcal{A}_{12}\{
[\hat{\boldsymbol{\mu}}_{1}\cdot  \hat{\boldsymbol{\mu}}_{2}
-(\hat{\boldsymbol{\mu}}_{1}\cdot
      \hat{\mathbf{r}}_{12})
(\hat{\boldsymbol{\mu}}_{2}\cdot
      \hat{\mathbf{r}}_{12})] \frac{\sin z}{z} +
[\hat{\boldsymbol{\mu}}_{1}\cdot  \hat{\boldsymbol{\mu}}_{2}
-3(\hat{\boldsymbol{\mu}}_{1}\cdot
      \hat{\mathbf{r}}_{12})
(\hat{\boldsymbol{\mu}}_{2}\cdot
      \hat{\mathbf{r}}_{12})] [\frac{\cos z}{z^{2}}-\frac{\sin z}{z^{3}}]\}$,
where $\mathcal{A}_{12}\equiv \frac{3\sqrt{\Gamma_1 \Gamma_2}}{8\pi}$.}.  $z\equiv nk_{0}r_{12}$; 
 $ \hat{\boldsymbol{\mu}}_{i}$, and $\hat{\mathbf{r}}_{12}$
are unit vectors along the transition dipole moments and along the relative separation between the molecules.
 
We stress  that Eq.~\eqref{master} is more general 
than that used in  \cite{yu2,bellomo} in the sense that it  accounts for qubit-qubit coupling, coherent external driving,  and nonlocal fluctuations due to radiative corrections \cite{agarwal}.

\section{Resonance fluorescence}

To evaluate the system's physical observables, we  perform the calculation of  the occupation probabilities in the steady state. We compute this for several different experimental conditions realizable in the laboratory, as shown in Fig. 1. These involve  weak and strong coherent  laser driving (e.g., Fig. 1(c)), different dipole  interaction strengths (Figs. 1(a), and 1(d)), and different   dimer's resonance frequencies (Fig. 1(b)). In particular,  we account for the fluorescence spectrum measured for the pair of coupled molecules reported in Ref.  \cite{hettich}. 

Our calculations are performed  in the standard two-qubit computational basis $\{\ket{00},\ket{01},\ket{10},\ket{11}\}$ ($\ket{ij}\equiv\ket{i_{1}}\otimes\ket{j_{2}}$), and the corresponding dimer density matrix elements are denoted as $\rho_{ij,kl}$, with associated populations $\rho_{ij,ij}$ (density matrix elements  $\ket{ij}\bra{ij}$),  $i,j=0,1$. The calculations here presented are amenable to experimental verification in a dimer  molecular   set-up  where the task of 
optically resolving the single molecules 
by means of  laser spectroscopy techniques is possible.
 This kind of experiment has been performed, for instance, by Hettich {\it et al}. \cite{hettich}, and  the detection and measurement of coupled single terrylene molecules embedded in an organic (para-terphenyl) crystal (thickness of about $250$~nm) has been reported. There the detection was carried out by means of fluorescence excitation spectroscopy at a temperature $\sim1.4$ K. Such a recorded  fluorescence spectra, at high excitation laser intensity, exhibits the appearance of an `additional'  peak  between those associated to the  individual transition frequencies  $\nu_1$, and $\nu_2$, which is the experimental evidence of the dipole-dipole coupling $V_{12}$ between the qubits. 

The continuous curves shown in  Fig. \ref{observable}(a) are in perfect agreement with the experimental results reported in \cite{hettich}, for the same set of implemented physical parameters: 
difference in dipole transition frequencies $\Delta_-=2320$ MHz, dipolar coupling strength $V_{12}=950$ MHz, individual and collective decay rates $\Gamma_1=\Gamma_2=2\pi\times 50$ MHz, $\Gamma_{12}=2\pi\times 9$ MHz; with $\Delta_e=-160$~MHz, for a laser coupling $\ell_i=200$~MHz. Here, we have calculated, directly from Eq. \eqref{master},  the steady state occupation probabilities for the states $\ket{01}$, $\ket{10}$, and $\ket{11}$ as a function of the laser detuning. 
We can also see from Fig. \ref{observable}(a), the effect due to a different dipolar coupling, as shown by the dashed curves, for $V_{12}=50$ MHz. Since this implies a change in the relative position of the coupled dipoles, this is reflected in a shift of the resonant peaks at frequencies $\nu_1$, and $\nu_2$, and a decreasing of the signal intensity  coming from the two-photon emission process (intermediate peak) due to the weaker dipolar coupling.

The case of molecules with equal transition frequencies ($\Delta_-=0$) is illustrated  by the  continuous curves in Fig. \ref{observable}(b), for $V_{12}=950$ MHz. As expected, only two peaks appear: one associated to the resonance  condition $\nu_1=\nu_2$, and the intermediate one, associated to the dipolar coupling. By bringing the dipoles into resonance, a clear increment  in the two-photon resonance signal, in comparison to the experimentally measured spectra \cite{hettich} (calculated as the dashed curve in Fig.  \ref{observable}(b)) is observed, thus indicating a stronger dipole-dipole coupling.

The resonance fluorescence spectrum reported in  \cite{hettich} 
can be   theoretically reproduced by plotting the steady state quantity   $\rho_{01,01}+\rho_{10,10}+2\rho_{11,11}$, as shown in Fig. \ref{observable}(c). 
This  shows the appearance of the two-photon resonance  as the laser excitation intensity is increased. This peak is shifted from the laser detuning  $\Delta_{+}=0$ due to the the vacuum fluctuation noise, an effective shift that has been experimentally \cite{hettich} and theoretically \cite{jh3}  estimated at around $-160$~MHz. The calculations here presented have been done for the independent and collective decay rates $\Gamma_i=18\pi$ MHz, $\Gamma_{12}=9\pi$ MHz.
 Figure \ref{observable}(d) shows a zoom of the intermediate resonance of  Fig. \ref{observable}(c), for a laser coupling  $\ell_i=100$ MHz, and different dipolar couplings.  This shows a clear increment in the size of the resonance as the qubit coupling is increased.
 If this graph were to be calculated for  $\ell_i=10$~MHz (not shown),  the probability of simultaneous excitation $\rho_{11,11}$ appears  multiplied by the  scale factor $10^{-5}$.
 Hence, the resonance signal can be tailored according to designed structural molecular parameters. The difference in the half-width of the reported peaks has its origin in the constructive and destructive  interference of the different  possible relaxation pathways \cite{Ficek}.

\begin{figure}[t]
  \centerline{\includegraphics[width=0.55\textwidth]{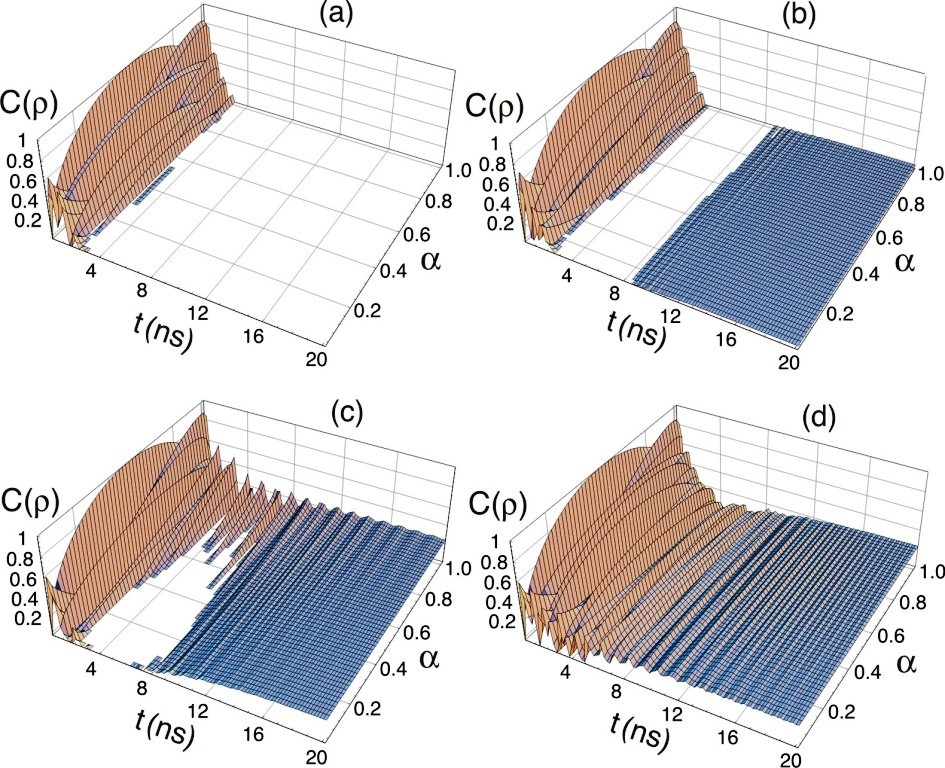}}
  \caption{Laser-driven entanglement dynamics control. 
  a) Early stage disentanglement:   
   $\Delta_+=0$, and $\ell_{1}=\ell_{2}=500$~MHz. b) The molecules couple to the laser with  different strengths: $\ell_{1}=300$~MHz,  
  and $\ell_{2}=500$~MHz. Figures c)  $\Delta_+=\Delta_{-}$, and d)  $\Delta_+=-\Delta_{-}$,  show the  difference in entanglement behavior when  the laser resonant condition is varied, for $\ell_{1}=300$~MHz, $\ell_{2}=500$~MHz. $\Delta_{e}=0$. 
In all  the graphs, $\Gamma_{1}=\Gamma_{2}=2\pi\times 50$~MHz, $\Gamma_{12}=2\pi\times 9$~MHz, $\Delta_-=2320$~MHz, and $V_{12}=950$~MHz.
 }
\label{fig2}
\end{figure}

\section{Dimer Entanglement Evolution}

As a norm for the entanglement quantification of the dipole-coupled molecules,  we use  Wootter's concurrence $
C(\rho)=max\{0, \lambda_1- \lambda_2- \lambda_3-\lambda_4\}$, where
the $\lambda_i$'s are the  square roots of the eigenvalues  of the 
non-hermitian matrix $\rho\widetilde{\rho}$, arranged in decreasing order, with 
$
  \widetilde{\rho}=\sigma^{(1)}_y \otimes \sigma^{(2)}_y 
  \rho^{\ast} \sigma^{(1)}_y \otimes \sigma^{(2)}_y 
$  \cite{wootter}.
We consider the initial state density matrix 
\begin{eqnarray}
\rho(0)=\left(
\begin{array}{cccc}
a  & 0 & 0 & w \\
0 & b & z & 0 \\
0 & z^{\ast} & c & 0 \\
w^{\ast} & 0& 0 & d
\end{array}\right) ,
\label{rho}
\end{eqnarray}
as  this allows the initialization of a broad class of entangled states \cite{yu2,yu4}. 
   In the standard two-qubit computational basis, the maximally  entangled  Bell state   $\ket{\psi^+}=\frac{1}{\sqrt{2}}(\ket{01}+\ket{10})$ is obtained for  
$a=w=d=0$; $b=c=z=1/2$. All the other states of the Bell basis, as well as the Werner mixed  state \cite{yu4} can be produced in a similar manner.

\subsection{Coherent control dynamics}

We consider  the dynamics of the initial entangled 
state  $\ket{\psi_{0}(\alpha)}=\sqrt{\alpha}\ket{01}+\sqrt{\beta}\ket{10}$, 
where $\alpha$ is a real number, $\sqrt{\beta}=|\sqrt{\beta}|e^{i\phi}$  is a complex number of  phase $\phi$, and 
$\alpha+|\beta|=1$.  This choice follows the identification of the   eigenstates of 
the system's Hamiltonian $\hat{H}_S$, in the absence of  external driving.
If no cooperative phenomena due to $V_{12}$ (decay rate $\Gamma_{12}$) take place (for example, due to a long  separation between the molecules, $V_{12}\sim 0$), in the absence of laser driving, 
the entanglement of this type of state has a natural asymptotic 
time decay due to the emission rates $\Gamma_{ii}\equiv \Gamma_{i}$ of the individual molecules. 

The numerical solution of the 
master equation \eqref{master} allows the quantification of the dimer entanglement  dynamics, as shown in Fig. \ref{fig2}, where the external field has been used to  produce  a coherent control of entanglement. 
Figure  \ref{fig2}(a) shows that for a strong laser coupling ($\ell_{i}=500$~MHz) 
the concurrence abruptly decays  to zero for a time less than $t_{\Gamma}$, where 
$t_{\Gamma}\equiv 1/\Gamma$ is the time associated to the decay rates   $\Gamma_{i}\equiv\Gamma$  for all the initial state configurations (all $\alpha$). In this work,   $\Delta_+$, $\Delta_-$, $\Delta_e$, $V_{12}$, and  $\ell_{i}$ are given in frequency ($\nu\equiv \omega/2\pi$) units, whereas   and $\Gamma_{ij}$  appear divided by  a $2\pi$-factor to consistently give the same units. 
This early stage disentanglement is due to the vacuum fluctuations affecting each molecule separately, and  radically 
changes when the coupling between the laser  and the molecules $\ell_1\neq\ell_2$, as shown in Fig. \ref{fig2}(b). There,   for all  $\alpha$, 
a finite, stationary entanglement $C(\rho)$ 
remains for long times. This behavior occurs after a period for which  the system experiences a complete disentanglement---$C(\rho)=0$  for {\it  all}  $\alpha$, and can be further controlled, as shown in Figs. \ref{fig2}(c) and (d),  by tuning the molecule resonance conditions with the laser field.

 In  Fig. \ref{fig2}(c),  
the ESD behavior is  partially suppressed  for finite $\alpha$ ($0.8\lesssim \alpha\leq 1$), when 
the laser is on resonance with the qubit 2 ($\nu_{L}=\nu_{2}$),  that is, for $\Delta_+=\Delta_{-}$ ($\nu_1>\nu_2$). In this case, the asymptotic limit of the concurrence is increased to $\sim 0.25$. In Fig \ref{fig2}(d),  the 
frequency of the external field has been tuned  to be in resonance 
with that of  qubit 1 
($\Delta_+=-\Delta_{-}$): the ESD behavior is now completely  suppressed  for all $\alpha$, and the entanglement shows an oscillatory decay. For long times, and for all  $\alpha$, the concurrence approaches the same stationary entanglement  as before.
 The fast oscillations of the concurrence are 
due to the competition between both the dipolar coupling and the laser-molecule  interactions. Thus, the results of  Figs. \ref{fig2}(c) and (d)  give a mechanism for coherent  control of  the dimer's ESD behavior, and for the  generation of  steady-state entanglement.

\begin{figure}[b]
  \centerline{\includegraphics[width=0.55\textwidth]{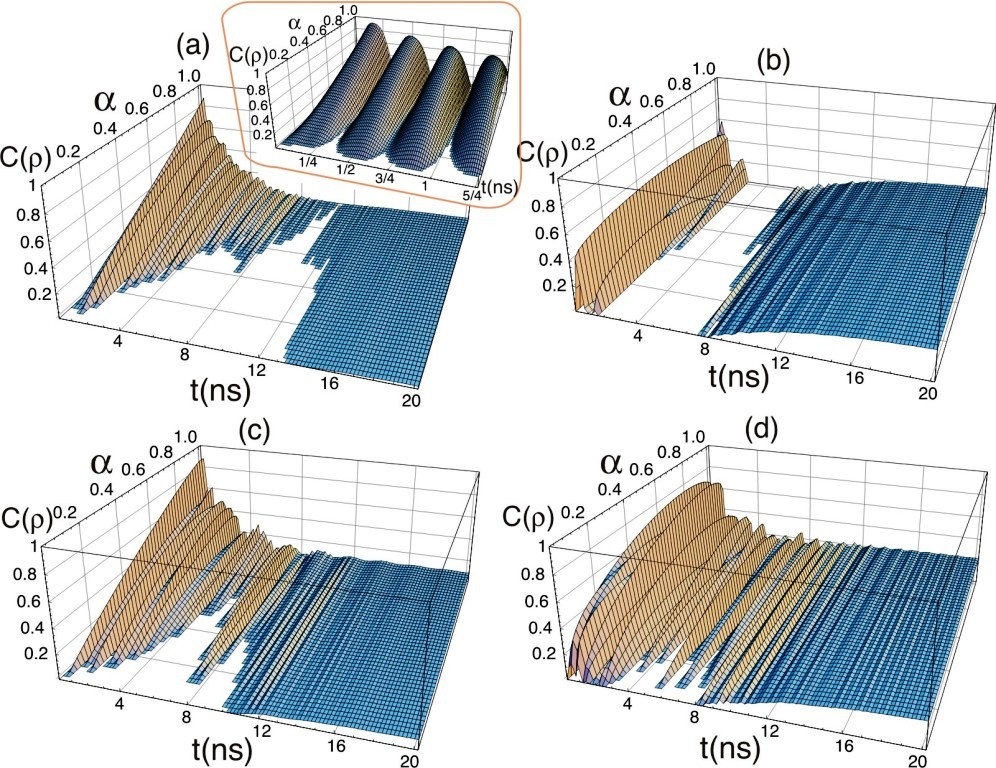}}
  \caption{Entanglement dynamics and ESB. Initial separable state $\ket{1}(\sqrt{\alpha}\ket{0}+\sqrt{\beta}\ket{1})$, (a) $\Delta_+=\Delta_{-}$,  $\ell_{1}=\ell_{2}=100$~MHz; (b) $\Delta_+=\Delta_{-}$, $\ell_{1}=300$~MHz,  $\ell_{2}=500$~MHz. (c) $\Delta_+=-\Delta_{-}$,  $\ell_{i}=100$~MHz; (d) $\Delta_+=-\Delta_{-}$, $\ell_{1}=300$~MHz, and $\ell_{2}=500$~MHz. $\Gamma_{1}=\Gamma_{2}=100\pi$~MHz, $\Gamma_{12}=18\pi$~MHz, $\Delta_-=2320$~MHz, $\Delta_e=-160$~MHz, and $V_{12}=950$~MHz.
 } 
\label{fig3f}
\end{figure}

Figure \ref{fig3f} shows another interesting behavior in the concurrence dynamics--the sudden birth of entanglement. In these graphs, the system is prepared in  the  initial  separable state $\ket{1_1}\otimes(\sqrt{\alpha}\ket{0_2}+\sqrt{\beta}\ket{1_2})$, and after a certain time, which depends on the type of input superposition (i.e., on the value of $\alpha$), entanglement appears in the system (see Figs. \ref{fig3f}(a), and (c)). Note that, due to the detuning between the molecular frequencies, the different resonant  laser tuning of the molecules produces different entanglement behavior. 

%
\begin{figure}[t]
  \centerline{\includegraphics[width=6.3cm]{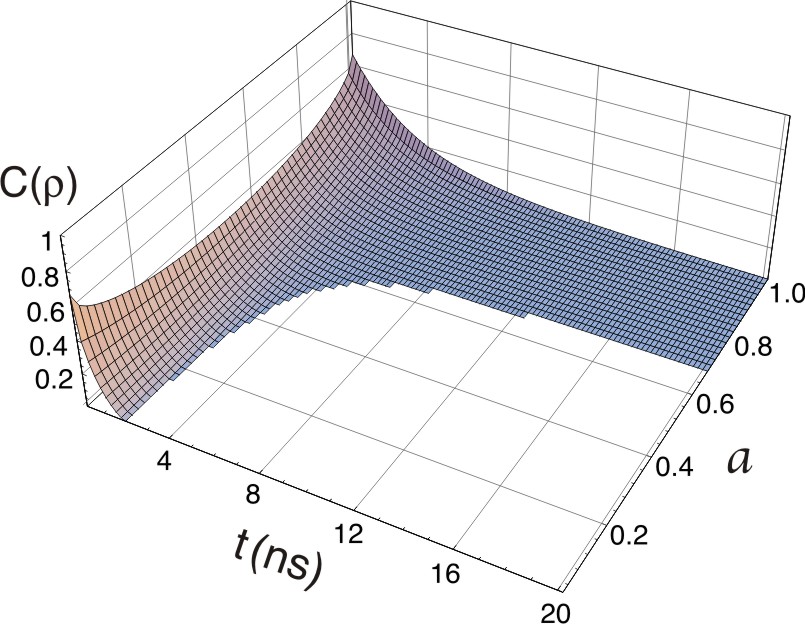}}
  \caption{Entanglement dynamics for  initial entangled states given by Eq.  
  \eqref{rho},  with $a\mapsto \frac{a}{3}$, $b=c=z=\frac{1}{3}$, 
$w=0$, and $d=\frac{1-a}{3}$; under  
 spontaneous emission. 
  $\Delta_-=2320$~MHz,  $\Delta_+/2=(\nu_{1}+\nu_{2})/2\approx 1\times 10^{9}$~MHz (optical transition). The decay rates $\Gamma_{i}\equiv \Gamma=100\pi$~MHz,   $\Gamma_{12}=0$; the remaining parameters ($V_{12}$, $\ell_{i}$ and $\Delta_{e}$) are set to zero.} 
\label{vacuum}
\end{figure}

Figure \ref{fig3f}(a) shows a delayed ESB for $\alpha=0$ and $\alpha$-values close to zero. The fast oscillations of the concurrence in Fig. \ref{fig3f}(a) are plotted in its inset for about $1/16$ of the total time-frame considered in the main figure. 
The graph  \ref{fig3f}(c) shows the effect due to the change in the laser resonance molecular tuning. Here,  $\Delta_+=-\Delta_{-}$, and   an appreciable `$\alpha$-region' of  early stage disentanglement  is  suppressed, accompanied by an increment in the amount of  generated stationary entanglement. 
As a major feature, we notice that due to the  laser addressing strategy (let's recall that the two dipoles have different transition energies), the initial states 
$\ket{0}\otimes \ket{1}$,  and $\ket{1}\otimes\ket{0}$ ($\alpha\equiv 1$) are more prone to a quick generation of  entangled states than the initial condition $\ket{1}\otimes \ket{1}$ ($\alpha\equiv 0$); in the latter there is an initial lack of entanglement generation, and ESB behavior takes place. In the former, the quick entanglement generation can be understood from the fact that a coherent superposition of the states 
$\ket{0}\otimes \ket{1}$  and $\ket{1}\otimes\ket{0}$ is naturally supported by the dipole-dipole interaction;  in fact, in the absence of laser tuning, such a superposition of states produce some of the eigenstates of the system's Hamiltonian.
The time scale for which ESB takes place depends on the qubit resonance conditions:  for the initial condition $\ket{1}\otimes \ket{1}$,   the ESB appearance in  Fig.  \ref{fig3f}(a) takes a longer time than that calculated for  Fig. 
  \ref{fig3f}(c).

Figures \ref{fig3f}(b) and (d) show the effect due to increasing the laser intensity. A  quicker generation of entanglement at around $\alpha \equiv 0$ is now produced (cf. Figs. \ref{fig3f}(a) and (c)), with a subsequent generation of ESD (Fig. \ref{fig3f}(b)), and a partial ESD suppression  (Fig. \ref{fig3f}(d)), followed by  an overall increment in the amount of generated  stationary entanglement, $C\sim 0.2$. These graphs manifest the effect due to the tuning of the molecular  resonance: if    $\Delta_+=-\Delta_{-}$ (Fig. \ref{fig3f}(d)) the time-frame for the region of zero concurrence of Fig. \ref{fig3f}(c) is almost completely suppressed, in addition of a larger number of entanglement oscillations before the steady-state is reached. For the chosen initial condition, the  $\Delta_+=-\Delta_{-}$-resonance condition favors  the entanglement production and inhibits ESD behavior. The same conclusion can be reached if we instead consider, as initial condition, the state $(\sqrt{\alpha}\ket{0}+\sqrt{\beta}\ket{1})\otimes\ket{1}$ (not shown).
The changes in the entanglement features shown in Fig.  \ref{fig3f}  are not only due to the chosen initial conditions and resonance features, but also to the fact that, in the considered scenarios, we have taken into account the effective energy shift $\Delta_{e}=-160$~MHz, which is  associated to the doubly excited state, as reported in ~\cite{hettich,jh3}. 
The numerical values  for the strength of the molecular interaction, the decay rates, and 
the  molecular transition energies used in the simulations of Fig. \ref{fig2} and   \ref{fig3f} are those reported in the experiment of Ref.~\cite{hettich}.

\begin{figure*}[t]
  \centerline{\includegraphics[width=0.98\textwidth]{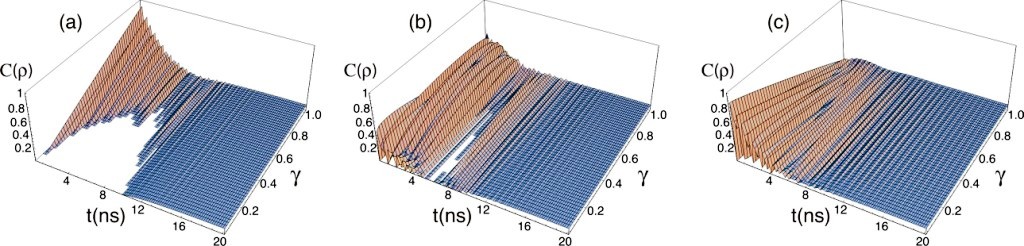}}
  \caption{Initial state preparation of  product states 
  $\ket{\Psi_0(\gamma)}=(\sqrt{\gamma}\ket{0}+\sqrt{\delta}\ket{1})\otimes(\sqrt{\zeta}\ket{0}+\sqrt{\eta}\ket{1})$,
  for  a) $\zeta=0$, b) $\zeta=1/2$, c) $\zeta=1$.  $\Delta_-=2320$~MHz,  $\Delta_+=0$. The decay rates $\Gamma_{i}=100\pi$~MHz,   $\Gamma_{12}=18\pi$~MHz.
   $\ell_{i}=100$~MHz,  $\Delta_{e}=-160$~MHz, and $V_{12}=950$~MHz.
  } 
\label{fig1+}
\end{figure*}

\subsection{Dependence on the initial state preparation}

Further to our previous analysis, we plot in Fig. \ref{vacuum} the entanglement 
dynamics for an initial state of the form Eq. \eqref{rho}, with $a\mapsto \frac{a}{3}$, $b=c=z=1/3$; 
$w=0$, and $d=\frac{1-a}{3}$.
Due to the structure of the density matrix, the concurrence allows an analytical calculation of the observed ESD behavior, and  the associated ESD time $\tilde{t}$  (for each $a$) can be exactly estimated. For example, for $a=0$,  $\tilde{t}=\frac{1}{\Gamma}\text{ln}\left| \frac{2+\sqrt{2}}{2}\right|$, which means  $\tilde{t}\approx 1.7$ ns $<\Gamma^{-1} $ in our molecular set-up. This behavior is  observed  over a continuous range of $a$ values, and for any $0\leq a < 2/3$ is given by 
\begin{equation}
 \tilde{t}= \frac{1}{\Gamma}\ln\left(\frac{1-a}{2-3a}\left[2-a+\sqrt{a^{2}-a+2}\right]\right).
   \label{tesda}
 \end{equation} 
For $a\geq 2/3$, ESD  is suppressed and entanglement decays in an asymptotic fashion. Hence, the dissipative molecular dynamics given by Eq.  \eqref{master} exhibit a rich behavior and  both an ESD dynamics and an asymptotic decay of entanglement can be induced in the dimer system by tuning the initial condition ($a$ value).

In Fig.  \ref{fig1+} we consider the entanglement dynamics for the general  initial product state of qubit superpositions 
\begin{equation}
\ket{\Psi_0(\gamma)}=(\sqrt{\gamma}\ket{0_1}+\sqrt{\delta}\ket{1_1})\otimes(\sqrt{\zeta}\ket{0_2}+\sqrt{\eta}\ket{1_2}),
  \label{ice}
\end{equation}
where, as before, $\gamma+|\delta|=1$, $\zeta +|\eta|=1$. Here,  we plot the corresponding concurrence as a function of $\gamma$, for   $\zeta=0,1/2,1$ ($\sqrt{\zeta}> 0$). We note that different initial preparations for the disentangled type of states lead to very different entanglement behavior. For instance, Fig.  \ref{fig1+}(a) exhibits a sudden birth of entanglement dynamics  for low values of $\gamma$, whereas for their intermediate values there are collapses and revivals that eventually become suppressed as $\gamma$ is increased; from this point, coherent oscillations of entanglement persist to finally decay and then converge to a stationary value.
In fact, the concurrence evolves, for all $\gamma$,  towards a finite non-zero stationary value due to the inter-qubit coupling strength  and the driving field. Also note that for short times the molecular entanglement reaches its maximum for $\gamma\mapsto 1$.  
Fig.  \ref{fig1+}(b) illustrates the case of equal superpositions in both qubits. Unlike  the previous case, here there is a finite entanglement oscillation at short times for  $\gamma\mapsto 0$, which can be understood in terms of the symmetry of the initial state. Here there is no presence of ESB behavior.  Due to the initial condition, Fig.  \ref{fig1+}(c) shows, in contrast to Fig.  \ref{fig1+}(a),  the highest entanglement oscillations for $\gamma \mapsto 0$.  
In all cases, the oscillations approach the same stationary value, as previously observed. Notice that in Fig.  \ref{fig1+} we have considered a non-zero value for $V_{12}$ and for the $\Delta_{e}$-shift, in contrast to the graph of Fig. \ref{vacuum}.
The parameters used in this simulation have been taken from the experimental data of Ref. \cite{hettich}. We have recently learned, although in a different context,  that steady state entanglement  is also produced if  counter-rotating wave terms are kept in the original system Hamiltonian \cite{scala,luis09b}.

\begin{figure*}[t]
  \centerline{\includegraphics[width=0.98\textwidth]{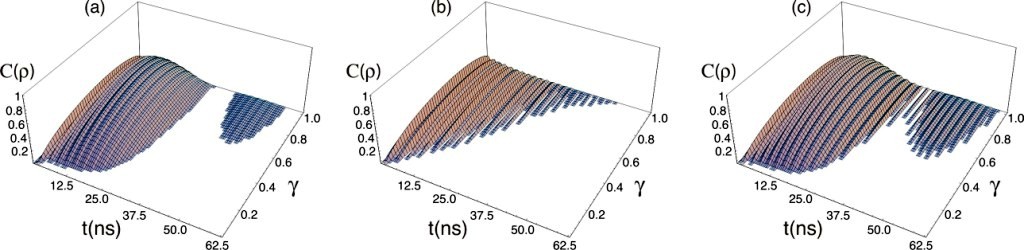}}
  \caption{Initial  product states $\ket{\Psi_{0}(\gamma)}$ (Eq. \eqref{ice}), 
  in the absence of qubit-qubit interaction and collective decay ($V_{12}=0$, $\Gamma_{12}=0$), for  a) $\zeta=0$, b) $\zeta=1/2$, c) $\zeta=1$.  $\Delta_-=2320$~MHz,  $\Delta_+=2638$~MHz.  $\Gamma_{i}=10\pi$~MHz, $\ell_{i}=200$~MHz,  $\Delta_{e}=-160$~MHz.
  }
\label{fig3+}
\end{figure*}

Figure \ref{fig3+} follows a similar  analysis as before, but for the case in which there is no physical interaction between the molecules, and therefore no collective decay. We consider instead the effects due to the vacuum fluctuations $\Delta_e$--the Lamb shift experienced by the molecular  doubly excited state, and  to the external  driving field. The results are clearly different from those of  Fig. \ref{fig1+}; even though the qubits start from  a separable state, they entangle due to the radiative corrections  and the external field. The concurrence exhibits a similar behavior for the conditions a) $\zeta=0$, and c) $\zeta=1$, for which there is a maximum amount of generated entanglement, and ESD and collapse-revival behavior are present as a function of the initial qubit preparation. The scenario $\zeta=1/2$ (graph  \ref{fig3+}(b)) does exhibit a different type of decay, and unlike the previous cases, there are not two ``islands" of entanglement, and the oscillations persist the most for $\gamma\mapsto 1$. Notice that the time scale for such oscillations 
is much longer than those of the graphs previously shown.

\subsubsection{Laser coupling}

\begin{figure}[h]
  \centerline{\includegraphics[width=0.55\textwidth]{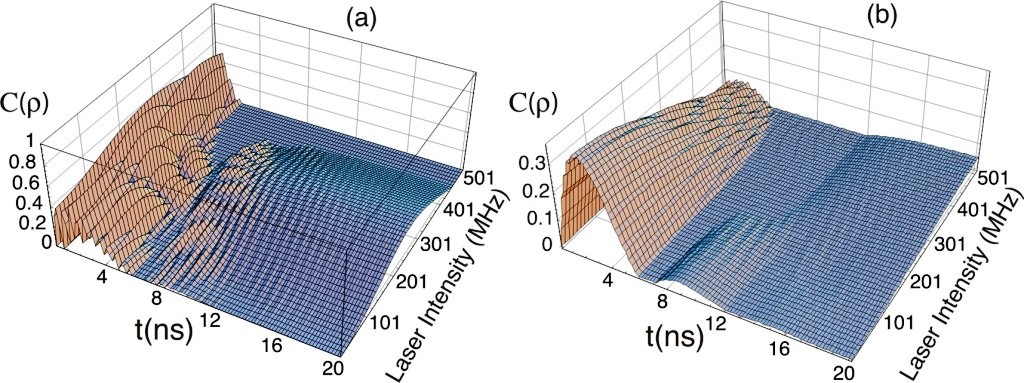}}
  \caption{Concurrence dynamics as a function of the laser intensity $\ell_i\equiv\ell$,  for the initial condition $\ket{\Psi_{0}(\gamma)}$, with  $\gamma=\zeta=1/2$. a) $V_{12}=950$~MHz, b) $V_{12}=50$~MHz.  $\Delta_-=2320$~MHz,  $\Delta_+=0$, $\Gamma=100\pi$~MHz,   $\Gamma_{12}=18\pi$~MHz,  $\Delta_{e}=-160$~MHz.
}
\label{fig5+}
\end{figure}

We address  the entanglement effects due to  the molecular coupling to the laser field, for the initial condition given by Eq. \eqref{ice}. Figure \ref{fig5+} shows the entanglement evolution  as a function of the laser field coupling strength $\ell$. In the figure, $\gamma=\zeta=1/2$, and the other parameters are given as in Fig.  \ref{fig1+}, for a) $V_{12}=950$ MHz, and b) $V_{12}=50$ MHz.
Figure \ref{fig5+}(a) shows that for the considered initial product state of superpositions there is an optimal laser coupling strength for which maximum stationary entanglement is generated. Away from this optimal value, the stationary  entanglement decreases monotonically until it reaches a certain intensity from which any entanglement correlation is washed away and the concurrence goes to zero after a few oscillations. This indicates that the laser can be used to tune and maximize stationary entanglement production, given an initial condition.
 
In Fig.  \ref{fig5+}(b) we have decreased $V_{12}$ to $50$~MHz. In contrast to the previous case, for low laser coupling strength, there are no entanglement oscillations before the concurrence reaches its minimum value: the concurrence exhibits an island whose height depends  on the laser strength. Unlike the previous case, here the maximum entanglement occurs at low intensity, due to the fact the qubit-qubit interaction is smaller than $\Delta_e$, which dominates the dynamics. The entanglement rapidly vanishes as the laser intensity increases, with  the two-particle state becoming  separable during  a period whose length increases with the laser intensity.  It then starts to  build up towards a stationary degree of entanglement, for which there is also a maximum value which is directly dependent on the   intensity of the laser. After this,  the entanglement gradually decays as the laser intensity is further increased (not shown) and eventually vanishes following a similar pattern to that of Fig. \ref{fig5+}(a).

\begin{figure}[h]
  \centerline{\includegraphics[width=0.5\textwidth]{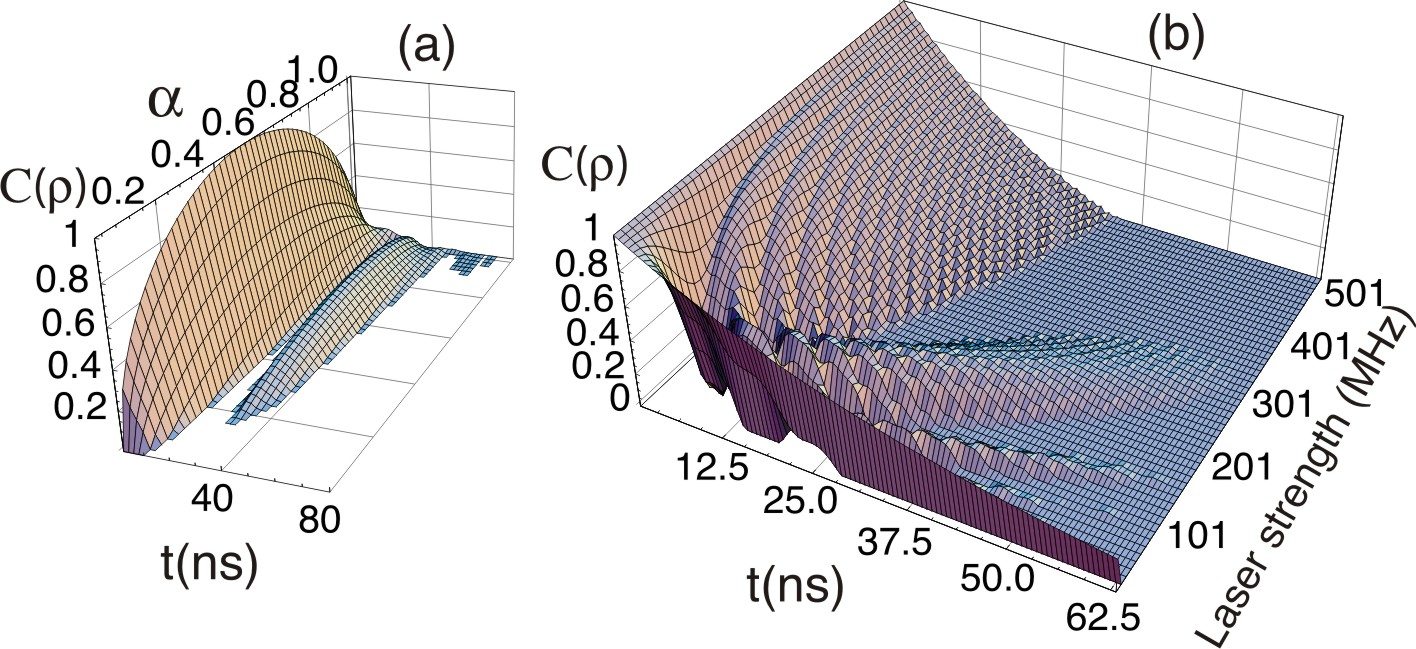}}
  \caption{Entanglement dynamics as a function of the initial condition and laser intensity, for the initial configuration $\ket{\psi_{0}(\alpha)}$.
  a) $\ell=200$~MHz, b) $\alpha=1/2$.  $\Delta_-=2320$~MHz,  $\Delta_+=2638$~MHz, $\Gamma=10\pi$~MHz,     $\Delta_{e}=-160$~MHz.
   $V_{12}=\Gamma_{12}=0$.
  }
\label{fig6+}
\end{figure}

For the sake of completeness, we next consider the situation of non-interacting qubits  ($V_{12}=0$), and no collective decay  for the initial state  $\ket{\psi_0(\alpha)}=\sqrt{\alpha}\ket{01}+\sqrt{\beta}\ket{10}$. 
Figure \ref{fig6+}(a) shows that ESD and collapse-revival behavior 
take place  in the entanglement evolution
as the initial condition is varied, by means of tuning $\alpha$ (e.g.,  maximal entanglement is attained for $\alpha=1/2$). In contrast, Fig. \ref{fig6+}(b) shows the dependence of entanglement on the laser intensity for the same initial  state   $\alpha=1/2$. Here, we distinguish three main dynamical scenarios. For low intensity values, there is an asymptotic decay of concurrence. As the intensity increases, collapse-revivals occur whereas for larger values of field coupling,  pure ESD phenomena take over the dynamics. The time scale plotted in Fig.  \ref{fig6+} is to be compared with the `natural' decay time $t_\Gamma\equiv \Gamma^{-1}$. It is clearly shown that 
coherent laser driving can be used as a key ingredient  for  determining the major features of entanglement dynamics.

\subsection{Nonlocal fluctuations: Lamb shift correction}

\begin{figure}[h]
  \centerline{\includegraphics[width=6.6cm]{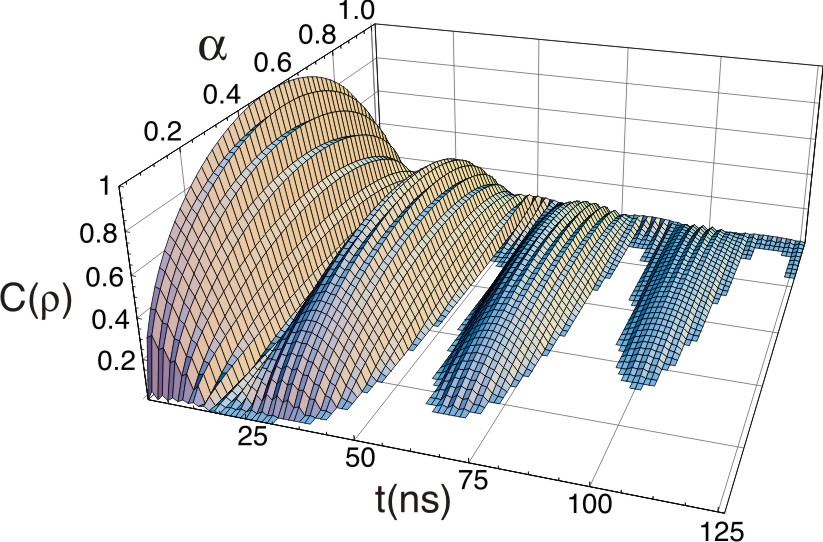}}
  \caption{Concurrence evolution  as a function of the  initial condition $\ket{\psi_{0}(\alpha)}$.
The qubits are  far apart from each other and do not experience a 
  collective decay ($V_{12}=0$, $\Gamma_{12}=0$). They, however,  
due to the vacuum noise, experience $\Delta_{e}=-160$~MHz ~\cite{hettich}. 
  $\Gamma_{i}=4\pi$~MHz, $\Delta_-=2320$~MHz,  $\Delta_+=2638$~MHz, and $\ell_{i}=200$~MHz.}
\label{fig3}
\end{figure}

We illustrate in Fig. \ref{fig3}  a major point of this work: in the scenario where the molecules are not subjected to the  collective damping 
due to the vacuum field, dark periods and revival of entanglement appear. This result is in contrast to
the revival of entanglement reported  in \cite{ficek1}, where such a behavior is critically dependent on the 
collective decay. As is shown in our case,  the revival phenomenon is strictly induced by the external laser control 
and the quantum noise fluctuations given by the Lamb shift \cite{jh3}.

The entanglement revival behavior 
occurs even when the pair of molecules do not interact, just as a consequence  of the 
noise due to vacuum electromagnetic fluctuations. Such 
fluctuations are  always present \cite{alicki02}, and the molecules experience a 
vacuum noise input and radiate a vacuum noise output. This  
effect shifts the energy levels of the molecules \cite{qnoise}. The  so-called  Lamb shift  has been  calculated for a specific molecular system 
in \cite{jh3}. We show such an effect in Fig. \ref{fig3}, where we have 
considered a  qubit separation such that $V_{12}=0$, and  $\Gamma_{12}=0$.  We 
have introduced a phenomenological $\Delta_{e}$-shift, starting from the experimental results reported in~\cite{hettich}, and the calculations of Ref. \cite{jh3}.

Our result is also in contrast with that obtained in 
Ref. \cite{bellomo}. There,  the authors  attribute the revival behavior (similar to ours) to 
the memory (non-Markovian) effects  induced by two  reservoirs acting on the qubits in an independent manner.
Figure \ref{fig3} shows  that the revival of entanglement is not, strictly speaking, a non-Markovian feature associated to an uncoupled two-qubit system.  This shows that for the case of   independent 
spontaneous emission decay,  a similar behavior to that reported in \cite{bellomo} can be found. Our result, however, has been derived  within a Born-Markov description, and as such,  does not invoke a   non-Markovian treatment. On the same grounds, this is 
different from  the collapse-revival behavior reported for the  interacting  system 
of  Ref.~\cite{sumanta}. The behavior showed in Fig. \ref{fig3} changes  with increasing emission rates (cf. Fig. \ref{fig6+}): when the  system decays more rapidly, the collapse-revival behavior disappears. On the other hand, Fig. \ref{fig6+}(b) clearly shows that, for a fixed rate,  collapse-revivals are suppressed for higher values of laser intensity, for which ESD becomes the relevant dynamical decay.

\begin{figure}[h]
  \centerline{\includegraphics[width=6.9cm]{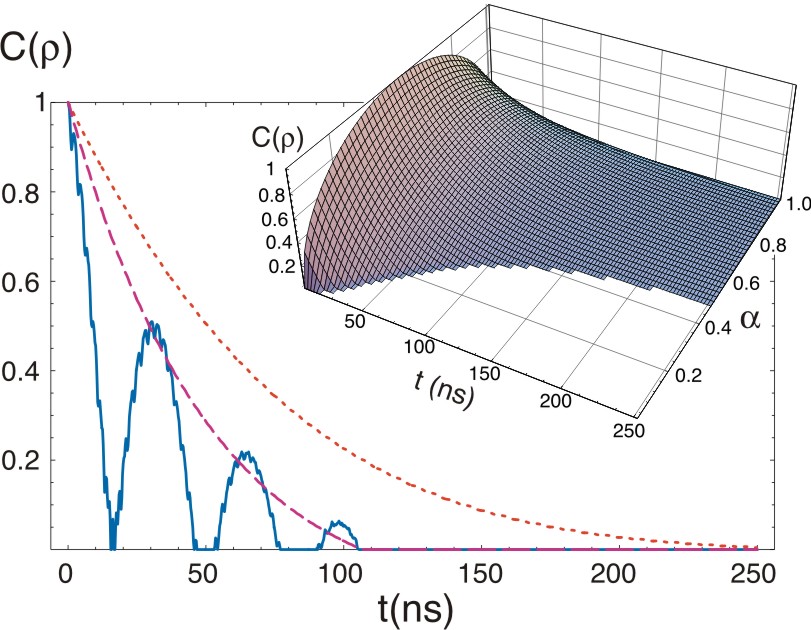}}
  \caption{Comparison between the effects due to  the  laser and the  
  vacuum fluctuations over the entanglement dynamics. In the main graph, the solid curve 
exhibits collapses and revivals; 
  $\ell_{i}=200$~MHz,
  $\Delta_{e}=-160$~MHz. The long dashed curve exhibits ESD;   $\ell_{i}=200$~MHz, $\Delta_{e}=0$. The short dashed curve corresponds to $\ell_{i}=1$~MHz, and $\Delta_{e}=-160$~MHz. The initial state is $\ket{\psi_{0}(1/2)}$. 
  The remaining parameters used in the simulations are the same as 
  in Fig. \ref{fig3}. By switching to the initial state $\sqrt{\alpha}\ket{00}+\sqrt{\beta}\ket{11}$, the inset shows the  ESD sensitivity  to the initial condition; $\Delta_+=20000$~MHz, $\Delta_{e}=-160$~MHz, and  $\ell_{i}=0$; the other parameters  are as in Fig. \ref{fig3}.}
\label{fig4}
\end{figure}

We have plotted,  in Fig. \ref{fig4},  the dynamics of the concurrence for 
the  initial entangled 
state  $\ket{\psi_{0}(\alpha=1/2)}$,   to show the interplay between the laser 
driving and the  shift  experienced by the doubly excited state due to background noise fluctuations. The concurrence decays asymptotically 
when the laser coupling is absent, in agreement with our previous results.

  The main graph of Fig. \ref{fig4} also 
shows sudden disentanglement (but not revival) when the laser coupling is 
strong and there is no contribution arising from the vacuum noise (long dashed curve). This behavior is also observed  in the inset of Fig. \ref{fig4}, for  a different 
 initial condition, $\sqrt{\alpha}\ket{00}+\sqrt{\beta}\ket{11}$: the ESD 
  is more pronounced when $\alpha$  tends to zero, and it vanishes  for $\alpha>1/2$. This is because in 
our density matrix configuration, $\rho_{00,00}$ corresponds to the lowest system
eigenstate in the standard basis, and hence, for $\alpha\rightarrow 1$, the system has a higher  probability of decaying to the ground state and the interesting quantum  effects are lost.

The sensitivity of the ESD dynamics  to the initial conditions is shown in  the inset of Fig.  \ref{fig4}.  In the main figure, the entanglement  decays   asymptotically (uppermost curve, $\ell=1$) for the initial conditions $\sqrt{\alpha}\ket{01}+\sqrt{\beta}\ket{10}$ (for {\it all} $\alpha$), and hence ESD never takes place. However, the inset of Fig.  \ref{fig4} shows that for the initial condition $\sqrt{\alpha}\ket{00}+\sqrt{\beta}\ket{11}$, and absence of laser driving, the entanglement decays abruptly for values of $\alpha$ close  to zero, and hence ESD is reinforced at very early times $\tilde{t}$. This clearly indicates that different types of initial entangled states experience different input noise from the environment. 

An analytical expression for the concurrence, for the molecular input values of the inset of Fig. \ref{fig4}, follows   from  the calculation of the  non trivial density matrix elements  found for  the solution of the master equation:
\begin{eqnarray}
\nonumber 
&& \rho_{00,00} = 1+(1-\alpha){\rm e}^{-2\Gamma t}\left(1-2{\rm e}^{\Gamma t}\right), \\ \nonumber 
&& \rho_{01,01} =\rho_{10,10}= (1-\alpha){\rm e}^{-2\Gamma t}\left(-1+{\rm e}^{\Gamma t}\right), \\ \nonumber
&& \rho_{11,11} = (1-\alpha){\rm e}^{-2\Gamma t}, \\ 
&& \rho_{00,11}=\rho_{11,00}^{\ast}=\sqrt{(1-\alpha)\alpha} \ {\rm e}^{\text{i} t(2\pi\Delta_{e}+\text{i}\Gamma+4\pi\Delta_{+})}.
\label{soluciondepsi}
\end{eqnarray}
The concurrence 
$C(\rho)=2\, {\text{max}}\, \{0,|\rho_{00,11}|-\rho_{01,01}\}$. Thus,  
the ESD time can be directly obtained from the condition  $|\rho_{00,11}|-\rho_{01,01}\leq 0$; for any  $\alpha<1/2$, this reads
\begin{equation}
\tilde{t}_\alpha=\frac{1}{\Gamma}\text{ln}\left| \frac{1-\alpha}{1-\alpha-\sqrt{\alpha-\alpha^2}}\right|
\label{tiempoesd} .
\end{equation}
For $\alpha\geq 1/2$ there exists no solution with physical meaning for $\tilde{t}_\alpha$, in agreement with what is shown in the inset of 
Fig. \ref{fig4}.
An experimental study 
on the engineering and modification  of the molecules decay rates has already been reported \cite{hettich2}; this could facilitate the observation of the effects reported in Figs. \ref{fig3} and \ref{fig4}. 

\section{conclusions}

We summarize the major findings of the present work on dimer entanglement dynamics. We have shown how the entanglement evolution, which includes ESD, ESB, and collapse-revival dynamics, can be precisely controlled by means of an external driving field, and an appropriate qubit initialization. In addition, the light-matter interaction can be used to produce transitions between the system states in such a way that a reliable production of steady state  entanglement is  favored. There are  entanglement dynamics corrections due to the radiative shift of the doubly-excited molecular state arising from  the vacuum fluctuations: in the scenario where the qubits do not experience collective coupling effects (only independent interaction with the environment), the entanglement exhibits a collapse-revival behavior. On the other hand, by considering different type of initial conditions, an ESD dynamics can also be induced due to the energy shift.  We have shown that it is crucial, for a correct characterization of entanglement dynamics,  to take into account the radiative corrections 
induced by the electromagnetic vacuum field, as such   shifts or energy corrections permanently affect the evolution  of entanglement.

\section{acknowledgements}

This work was partially supported by Colciencias Grant
1106-452-21296, and the  DAAD-Colciencias exchange (PROCOL)  program. We acknowledge the hospitality of the
Freiburg Institute for Advanced Studies, and the
Max Planck Institute for the Science of Light Erlangen,
 where parts of this work were completed. JHR is grateful to the DAAD for a Research Stay Award (A/09/0851). We would like to acknowledge L. L. S\'anchez-Soto for a critical reading of the manuscript.

\end{document}